\documentclass[twocolumn,english,aps,reprint,groupedaddress,notitlepage,nobibnotes,nofootinbib,preprintnumbers,showpacs]{revtex4-1}
\usepackage{lmodern}

\usepackage[T1]{fontenc}
\usepackage[latin9]{inputenc}
\usepackage{geometry}
\usepackage{color}
\usepackage{babel}
\usepackage{mathtools}
\usepackage{amsmath}
\usepackage{amssymb}
\usepackage{graphicx}
\usepackage{esint}
\usepackage[unicode=true,pdfusetitle,
 bookmarks=true,bookmarksnumbered=false,bookmarksopen=false,
 breaklinks=false,pdfborder={0 0 1},backref=false,colorlinks=true]
 {hyperref}
\hypersetup{
 citecolor=black,linkcolor=black,urlcolor=black}
\makeatletter

 \@ifundefined{textcolor}{}
 {
   \definecolor{BLACK}{gray}{0}
   \definecolor{WHITE}{gray}{1}
   \definecolor{RED}{rgb}{1,0,0}
   \definecolor{GREEN}{rgb}{0,1,0}
   \definecolor{BLUE}{rgb}{0,0,1}
   \definecolor{CYAN}{cmyk}{1,0,0,0}
   \definecolor{MAGENTA}{cmyk}{0,1,0,0}
   \definecolor{YELLOW}{cmyk}{0,0,1,0}
 }
\usepackage{babel}
\makeatletter
\def\simgt{\mathrel{\lower2.5pt\vbox{\lineskip=0pt\baselineskip=0pt
           \hbox{$>$}\hbox{$\sim$}}}}
\def\simlt{\mathrel{\lower2.5pt\vbox{\lineskip=0pt\baselineskip=0pt
           \hbox{$<$}\hbox{$\sim$}}}}
\makeatother

\newcommand{\be}{\begin{equation}}
\newcommand{\ee}{\end{equation}}
\newcommand{\bea}{\begin{eqnarray}}
\newcommand{\eea}{\end{eqnarray}}
\newcommand{\eq}[1]{\begin{align}#1\end{align}}

\newcommand{\Eq}[1]{Eq.~\eqref{#1}}

\begin{document}

\preprint{\hbox{CALT-TH-2016-035} }

\title{Symmetry for Flavor-Kinematics Duality from an Action}

\author{Clifford Cheung and Chia-Hsien Shen}
\affiliation{Walter Burke Institute for Theoretical Physics \\California Institute of Technology, Pasadena, CA 91125}
\date{\today}
\email{clifford.cheung@caltech.edu, chshen@caltech.edu}
\pacs{}
\begin{abstract}

We propose a new representation of the nonlinear sigma model that exhibits a manifest duality between flavor and kinematics. The fields couple exclusively through cubic Feynman vertices which define the structure constants of an underlying kinematic algebra.   The action is invariant under a combination of internal and spacetime symmetries whose conservation equations imply flavor-kinematics duality, ensuring that all Feynman diagrams satisfy kinematic Jacobi identities. Substituting flavor for kinematics, we derive a new cubic action for the special Galileon theory.  In this picture, the vanishing soft behavior of amplitudes is a byproduct of the Weinberg soft theorem.


\end{abstract}

\maketitle

\section{Introduction}

The study of scattering amplitudes has uncovered a beautiful duality between gauge theory and gravity concisely  summarized by the mantra, gravity $\sim$ gauge$^2$.   This connection was first understood for tree amplitudes in Yang-Mills (YM) theory and gravity in the context of open-closed string duality \cite{Kawai:1985xq}. Much later, Bern, Carrasco, and Johansson (BCJ)~\cite{Bern:2008qj} proposed a remarkable generalization of this squaring relation known as color-kinematics duality.  BCJ showed that tree amplitudes in YM theory can be rearranged into the  schematic form,
\eq{
A \sim \sum_i \frac{C_i N_i}{D_i},
\label{eq:colorAmp}
}
where $i$ sums over cubic topologies with propagator denominators $D_i$, color structures $C_i$, and kinematic numerators $N_i$.   Here $C_i$ and $N_i$ satisfy Jacobi identities,
\eq{
C_i + C_j + C_k = 0  \quad \textrm{and} \quad
N_i + N_j + N_k = 0 ,
\label{eq:Jacobi}
}
where $i,j,k$ denote any triplet of cubic topologies which are the same except for a single propagator.
That there exist $N_i$ with the same algebraic relations as $C_i$ is at the heart of color-kinematics duality.  Since $N_i$ and $C_i$ are in this sense interchangeable, we can substitute the latter with the former, yielding the double copy,
\eq{
M \sim \sum_i \frac{N_i N_i}{D_i},
\label{eq:gravAmp}
}
which is the graviton tree amplitude~\cite{Bern:2010ue,Bern:2010yg}.
The double copy has been generalized to include loops, supersymmetry, and matter fields (cf. \cite{Carrasco:2015iwa} and references therein).

While color Jacobi identities are trivialized by an underlying Lie algebra, this is not so simple for kinematics. BCJ strongly suggests an underlying algebra for kinematic numerators, but this structure remains elusive except in limited contexts, {\it e.g.}~for YM in the self-dual sector \cite{Monteiro:2011pc} and in the formalism of Cachazo, He, and Yuan (CHY) \cite{Cachazo:2013gna,Cachazo:2013hca,Cachazo:2013iea,Monteiro:2013rya}.  

In constrast, the nonlinear sigma model (NLSM) is a theory of Nambu-Goldstone bosons unburdened by gauge symmetry, thus offering a simpler path to the kinematic algebra.  Flavor-kinematics duality in the NLSM has been explored at tree-level and on the worldsheet \cite{Chen:2013fya, Du:2016tbc,Carrasco:2016ldy}, though without mention of the double copy.

In this letter, we present a new formulation of the NLSM in general spacetime dimension.  
This action is remarkably simple, comprised of just a handful of fields interacting via cubic Feynman vertices that play the role of structure constants in an underlying kinematic algebra.  Flavor-kinematics duality then emerges as a  symmetry: the kinematic Jacobi identities are literally the current conservation equations for a certain combination of internal and spacetime symmetries.  In turn, all Feynman diagrams automatically satisfy \Eq{eq:Jacobi}.  Applying the double copy construction, we then derive a new cubic action for the special Galileon theory~\cite{Cachazo:2014xea,Cheung:2014dqa,Hinterbichler:2015pqa}, which describes a scalar coupled through a tower of higher derivative interactions.  
Lastly, we show how these formulations reproduce the vanishing soft behavior of amplitudes.


\section{Warmup}

\label{eq:biadjoint}

As a preface to our main results, let us briefly review the theory of a biadjoint scalar.  Though trivial in structure, this theory nicely illustrates how Jacobi identities arise from considerations of symmetry.  The action is
\eq{
S = \int \frac{1}{2}\phi^{a\bar a} \Box \phi^{a\bar a} + \frac{\lambda}{3} f^{abc}  f^{\bar a \bar b \bar c} \phi^{a \bar a} \phi^{b \bar b} \phi^{c \bar c},
}
where $f^{abc}$ and $f^{\bar a\bar b\bar c}$ are the structure constants for a pair of global flavor symmetries.
The equations of motion are
\eq{
\frac{\delta S}{\delta \phi^{a \bar a}} = \Box \phi^{a \bar a} +  \lambda f^{abc}  f^{\bar a \bar b \bar c}  \phi^{b \bar b} \phi^{c \bar c}.
\label{eq:biadjEOM}
}
The action is invariant under the global flavor rotations,
\eq{
\delta \phi^{a \bar a} &=  f^{abc} \theta^b \phi^{c \bar a} ,
}
whose associated Noether current is
\eq{
J^{a}_\mu &= - f^{abc} \phi^{b \bar a} \partial_\mu \phi^{c \bar a}.
}
Noether current conservation then implies that
\eq{
&\partial J^{d} = - f^{dae} \phi^{a \bar a}  \overset{\leftrightarrow}{\Box} \phi^{e \bar a} = 0 \label{eq:color_Jacobi} \\
&= \frac{\lambda}{3} f^{\bar{a}\bar{b}\bar{c}}  \phi^{a\bar{a}} \phi^{b\bar{b}}\phi^{c\bar{c}} \left(
	f^{dae}f^{ebc} + f^{dbe}f^{eca}+ f^{dce}f^{eab} \right), \nonumber
}
which is the Jacobi identity. Here $\smash{\overset{\leftrightarrow}{\Box}=\frac{1}{2}(\overset{\rightarrow}{\Box}-\overset{\leftarrow}{\Box})}$ and we have used \Eq{eq:biadjEOM} together with the cyclicity of $f^{\bar a \bar b \bar c}$.

The above derivation is actually equivalent to the diagrammatic representation of the Jacobi identity that typically appears in the study of scattering amplitudes.  In terms of Feynman diagrams, the d'Alembertian in \Eq{eq:color_Jacobi} has the effect of canceling an internal propagator.  The resulting triplet of objects---each equal to the Feynman diagrammatic numerator associated with a given cubic topology---satisfies the Jacobi identity.


\section{Action}
\label{sec:setup}

The strategy above can be applied directly to the NLSM, though doing so requires an alternative formulation of the theory.  To begin, let us introduce the fields
\eq{
X_\mu ,\quad Y ,\quad Z^{\mu},
}
 in the adjoint representation of a flavor symmetry.  The $XYZ$ fields interact via the remarkably simple action
\eq{
		S &= 
 \int Z^{a\mu}  \boxdot X^{a}_{\mu}+    \frac{1}{2} \, Y^a  \boxdot Y^{a} ,
		\label{eq:action}
}
where we have defined a modified d'Alembertian,
\eq{
\boxdot ()^a&= \Box ()^a  + 2 f^{abc} Z^{b\mu} \partial_\mu ()^c.
}
Expanded fully, the action becomes
\eq{
S =  \int& 
\;\, Z^{a\mu}  \Box X^{a}_{\mu}  +\frac{1}{2}Y^a  \Box Y^{a} \nonumber \\
 &-    f^{abc}\left(    Z^{a\mu}  Z^{b\nu} X^c_{\mu\nu} 
+     Z^{a\mu} ({Y}^b \overset{\leftrightarrow}{\partial_\mu} {Y}^{c}) \right)
\label{eq:actionexpand}, 
} 
where $X_{\mu\nu} = \partial_\mu X_\nu - \partial_\nu X_\mu$ and $\smash{\overset{\leftrightarrow}{\partial_\mu} = \frac{1}{2} (\overset{\rightarrow}{\partial_\mu}-\overset{\leftarrow}{\partial_\mu})}$.  
As we will soon see, the Nambu-Goldstone bosons of the NLSM are simultaneously described by $Y$ and $Z$, so this formulation obscures Bose symmetry.   Moreover, the cubic structure of the action hides the underlying parity of the NLSM interactions.  These properties are the cost of manifesting flavor-kinematics duality as a symmetry.

\section{Scattering Amplitudes}

The structure of the action in \Eq{eq:action} is reminiscent of a ``colored scalar'', $Y$, coupled to the $(-)$ and $(+)$ components of a ``YM field'', $X$ and $Z$.  The interactions in \Eq{eq:actionexpand} are then analogous to cubic $\overline{\rm MHV}$ vertices.  By general arguments \cite{Cheung:2015aba}, all tree amplitudes trivially vanish except those with exactly two $Y$ states or exactly one $X$ state, with all other states given by $Z$.

Our claim is that the tree amplitudes of the NLSM are equal to the tree amplitudes
\eq{
A(\ldots, Y_i, \ldots, Y_j, \ldots)\label{eq:amp},
}
where $i,j$ are arbitrary and the ellipses denote all other external particles, taken to be longitudinally polarized $Z$ states for which $\epsilon_\mu = i k_\mu$ in units of the NLSM decay constant. Note that Bose symmetry is ultimately preserved since the final amplitude does not depend on which particles are chosen to be $Y$ states.   

As an illustration of this, let us turn to the four-particle amplitude.
Using Feynman diagrams, we compute the kinematic numerators for the half-ladder topology for $(Y_1, Z_2, Z_3, Y_4)$, yielding
\eq{
N_s = s^2, \quad
N_t = s^2 - u^2, \quad
N_u = u^2,
}
where $N_t = N_s -N_u$ so the kinematic Jacobi identity is satisfied.  Moreover, the resulting flavor-ordered amplitude precisely matches that of the NLSM,
\eq{
A_4 = \frac{1}{2}(s+t),
}
in the convention that $[T^a, T^b] = i\sqrt{2}  f^{abc} T^c$.
Squaring the numerators via the double copy procedure, we obtain
\eq{
M_4 = - stu,
}
which is the amplitude of the special Galileon.

Alternatively, we could have instead computed the kinematic numerators for the choice $(Y_1,Y_2, Z_3, Z_4)$,
\eq{
N_s = t^2-u^2, \quad
N_t = t^2, \quad
N_u = - u^2,
}
or for the choice $(Y_1,Z_2, Y_3, Z_4)$,
\eq{
N_s = - s^2, \quad
N_t =  - t^2, \quad
N_u = t^2 - s^2 ,
}
which give different numerators but the same amplitude.

We can generalize to $n$-particle scattering by computing all the kinematic numerators for half-ladder topologies, which form a complete basis for all tree amplitudes~\cite{Bern:2008qj,DelDuca:1999rs}.  For later convenience, we define 
\eq{
\tau_i = -2 p_i \sum_{j<i} p_j \quad \textrm{and} \quad
(\tau_i^\pm)_\mu^{\;\;\nu} =  \delta_\mu^\nu \tau_i \pm 2 p_{i\mu} p_i^\nu,
}
as well as the kinematic variables
\eq{
\Sigma_{ij} &= \tau_i \tau_{i+1} \ldots \tau_{j-1} \tau_j \nonumber \\
\Sigma_{ij}^\pm &= 2p_i \tau^\pm_{i+1} \tau^\pm_{i+2} \ldots  \tau^\pm_{j-2} \tau^\pm_{j-1} p_j.
}
For each choice of $Y$ states, it is a straightforward exercise to calculate the corresponding half-ladder numerators via Feyman diagrams, yielding
\eq{
N(Y_1, \ldots, Y_{n}) &= -\Sigma_{2,n-1} \nonumber \\
N(Y_1, \ldots, Y_{i},\ldots) &= \Sigma_{2,i-1} \Sigma^-_{i,n} \nonumber \\
N(\ldots, Y_i, \ldots, Y_{n}) &= - \Sigma^+_{1,i} \Sigma_{i+1,n-1} \nonumber \\
N(\ldots, Y_i, \ldots, Y_j, \ldots) &=  \Sigma^+_{1,i} \Sigma_{i+1,j-1} \Sigma^-_{j,n} ,
\label{eq:numerators}
}
where the ellipses denote external $Z$ states.  
The first line of \Eq{eq:numerators} is the simple numerator proposed in \cite{Carrasco:2016ldy,Carrasco:2016ygv}.
We have checked that these expressions reproduce the tree amplitudes of the NLSM up to ten-particle scattering.

\section{Equations of Motion}

With the help of Feynman diagrams it is simple to check flavor-kinematics duality in specific examples.  However, to derive more general principles, it will be convenient to study the classical field equations, which are a proxy to tree-level Feynman diagrams \cite{Schwartz:2013pla}.  The Euler-Lagrange equations of motion for \Eq{eq:action} are
\eq{
		\frac{\delta S}{\delta X^{a}_{\mu}} &= \Box Z^{a \mu} + f^{abc}  (2Z^{b \nu} \partial_\nu Z^{c \mu}+2\partial_\nu Z^{b\nu}  Z^{c \mu}) =0  \nonumber \\
		\frac{\delta S}{\delta Y^a} &= \Box Y^a + f^{abc} (2 Z^{b\nu} \partial_\nu Y^{c} + \partial_\nu Z^{b \nu} Y^c)   =0 \nonumber \\
		\frac{\delta S}{\delta Z^{a\mu}} &= \Box X^{a}_{\mu} - f^{abc} 
		\left(  2 Z^{b \nu}  X^{c}_{\mu\nu} + Y^{b}\partial_{\mu} Y^{c}\right) =0, \label{eq:EOM}
}
where the divergence of the first equation yields
$\Box \partial_\mu Z^{ \mu} = 0$.  This implies the classical field condition
\eq{
\partial_\mu Z^{ \mu} = 0,
\label{eq:dZ}
}
whenever $Z$ is an off-shell source. If $Z$ is on-shell, then by the prescription of \Eq{eq:amp} it is longitudinal polarized, so \Eq{eq:dZ} is still valid because of the on-shell condition.  In any case, the bottom line is that \Eq{eq:dZ} is generally true whenever the equations of motion are satisfied.

Since \Eq{eq:dZ} is a constraint on classical fields, its implications for amplitudes are actually somewhat subtle.  In particular, due to the nondiagonal kinetic term, 
an off-shell source $Z$ propagates into $X$, which can then only interact via the field strength combination in \Eq{eq:actionexpand}.  From this perspective \Eq{eq:dZ} simply says that the longitudinal polarizations of $X$ are projected out.



\section{Symmetries}

The action in \Eq{eq:action} has a surprisingly rich set of local and global symmetries. We now present these symmetries and derive their associated Noether currents.

\smallskip
\noindent {\bf Local Transformation}
\smallskip

\noindent To begin, consider the local transformation,
\eq{
\delta X_\mu =  \partial_\mu \theta,
}
for an adjoint-valued gauge parameter $\theta$.
Modulo boundary terms, the action shifts by
\eq{
\delta S   =  - \int   \partial_\mu  Z^{a \mu} \Box \theta^a,
}
which is zero on the equations of motion by \Eq{eq:dZ}.  

\smallskip
\noindent {\bf Global $\boldsymbol{\delta_X}$ Transformation}
\smallskip

\noindent The first global symmetry transformation is
\eq{
\delta_X X_\mu &= \theta_{X\mu \nu} Z^\nu \nonumber  \\
\delta_X Y &= 0  \nonumber \\
\delta_X Z^\mu &= 0,
\label{eq:deltaX}
}
where $\theta_{X \mu\nu} = \partial_\mu \theta_{X\nu} -\partial_\nu \theta_{X\mu}$ is a constant antisymmetric matrix.  
While $\theta_{X\mu}$ is technically spacetime-dependent, it always enters with a derivative so the symmetry is still global.  The action shifts by
\eq{
\delta_X S  = \theta_{X\mu\nu}  \int f^{abc} \partial_\rho  Z^{a \rho} Z^{b\mu} Z^{c \nu},
}
which again vanishes on the equations of motion.

\smallskip
\noindent {\bf Global $\boldsymbol{\delta_Y}$ Transformation}
\smallskip

\noindent The second global symmetry  transformation is
\eq{
\delta_Y X_\mu &=  \theta_{Y\mu } Y  \nonumber \\
\delta_Y Y &=  -  \theta_{Y\mu} Z^\mu  \nonumber \\
\delta_Y Z^\mu &= 0, \label{eq:deltaY}
}
where $\theta_{Y\mu} = \partial_\mu \theta_Y$ is a constant vector.   Again, $\theta_Y$ is spacetime-dependent but the symmetry is still global.  The action transforms as 
\eq{
\delta_Y S  = \theta_{Y\mu}  \int f^{abc} \partial_\rho  Z^{a \rho} Z^{b\mu} Y^{c},
}
which is zero on the equations of motion.

\smallskip
\noindent {\bf Global $\boldsymbol{\delta_Z}$ Transformation}
\smallskip

\noindent The third global transformation is
\eq{
\delta_Z X_\mu &=\theta_{Z}^{\nu} \partial_\nu X_\mu +   \partial_\mu \theta_{Z}^{\nu} X_\nu  \nonumber \\
\delta_Z Y &=  \theta_{Z}^{\nu} \partial_\nu Y  \nonumber \\
\delta_Z Z^\mu &= \theta_{Z}^{\nu} \partial_\nu Z^\mu  -  \partial_\nu \theta_{Z}^{\mu} Z^\nu \label{eq:deltaZ},
}
for a transverse parameter, $\partial_\mu \theta_Z^\mu =0$.
This transformation is an infinitesmal diffeomorphism, 
where $Y$ transforms as a scalar and $X$ and $Z$ as vectors.  Here we will restrict to Poincare transformations, 
$\theta_Z^\mu =  a^\mu + b^\mu_{\;\;\nu} x^\nu$,
where $a$ is a constant translation vector and $b$ is a constant antisymmetric rotation matrix.

\smallskip
\noindent {\bf Noether Current Conservation}
\smallskip

\noindent The global transformations $\delta_X$, $\delta_Y$, $\delta_Z$ are associated with a set of equations for Noether current conservation,
\eq{
-\theta_{X\mu} \partial J^{\mu}_X &= \delta X_\mu^a  \frac{\delta S}{\delta X_\mu^a} \nonumber \\
-\theta_{Y} \partial J_Y &= \delta X_\mu^a  \frac{\delta S}{\delta X_\mu^a}+ \delta Y^a  \frac{\delta S}{\delta Y^a} \nonumber \\
-\theta_{Z}^{\mu} \partial J_{Z\mu} &= \delta X_\mu^a  \frac{\delta S}{\delta X_\mu^a}+ \delta Y^a  \frac{\delta S}{\delta Y^a} + \delta Z^{a\mu}  \frac{\delta S}{\delta Z^{a\mu}} ,
}
which with the transverse condition on $\theta_Z$ imply
\eq{
\partial J_X^{\mu} &=	- 2Z^{a\nu} \overset{\leftrightarrow}{\Box} \partial_\nu Z^{a\mu}  \nonumber \\
	\partial J_Y &= 	- 2Z^{a\nu} \overset{\leftrightarrow}{\Box} \partial_\nu Y^a  \nonumber \\
	\partial J_{Z\mu} &=2Z^{a\nu}  \overset{\leftrightarrow}{\Box} X_{\mu \nu}^a +Y^a  \overset{\leftrightarrow}{\Box} \partial_\mu Y^a,
}
which is the analog of \Eq{eq:color_Jacobi} for the NLSM.  Note that the cubic interactions also happen to be invariant under the local versions of the $\delta_X$, $\delta_Y$, $\delta_Z$ transformations.

\section{Kinematic Algebra}

To derive the kinematic algebra it is useful to introduce a unified description of the $XYZ$ fields in terms of an adjoint-valued multiplet, 
\eq{
W_A &= \left[ \begin{array}{c}
	X_\mu \\
	Y \\
	Z^{\mu}  \end{array} \right] ,
}
so the action in \Eq{eq:actionexpand} becomes
\eq{
S &= \int  \frac{1}{2} g^{AB} W^{a}_A \Box W^{a}_B +\frac{1}{3} f^{abc}F^{ABC} W^a_A W^b_B W^c_C.
}
Here capital Latin indices are raised and lowered by
\eq{
g_{AB} = \left[ \begin{array}{ccc}
0 & 0 & \delta_{\nu}^\mu \\
0 & 1  & 0 \\
 \delta_{\mu}^{\nu} & 0 & 0\end{array} \right] \;\;\; \textrm{and} \;\;\; g^{AB} = \left[ \begin{array}{ccc}
0 & 0 & \delta_{\mu}^{\nu} \\
0 & 1  & 0 \\
\delta_{\nu}^\mu & 0 & 0\end{array} \right] .
\label{eq:metric}
}
The kinematic structure constant $F^{ABC}$ is a differential operator acting multilinearly on the fields, and it describes the cubic Feynman vertex.  Contracted with a single field, $F^{ABC}W_B$ is represented by a matrix,
\eq{
 \left[ \begin{array}{ccc}
	0  & 0 &   -\delta^\mu_\nu  \overset{\leftarrow}{\partial} Z  + \overset{\leftarrow}{\partial_\nu} Z^{\mu}  \\
	0 &  \frac{1}{2} (Z \overset{\rightarrow}{\partial} -   \overset{\leftarrow}{\partial} Z) &    \frac{1}{2} ( \overset{\leftarrow}{\partial_\nu} Y  - \partial_\nu Y)\\
	\delta_{\mu}^{\nu} Z \overset{\rightarrow}{\partial}  - Z^{\nu}  \overset{\rightarrow}{\partial_\mu} & \frac{1}{2}( -  Y \overset{\rightarrow}{\partial_\mu} + \partial_\mu Y) &  \partial_\mu X_{\nu} -\partial_\nu X_{\mu} 
 \end{array} \right] ,
 \label{eq:generator}
}
which is manifestly antisymmetric as required.  The commutative subgroup of the $\delta_Z$ transformations, {\it i.e.}, spacetime translations, form a natural Cartan subalgebra.  In turn, the root vectors are literally momenta while $\delta_X$ and $\delta_Y$ are raising and lowering operators.  We leave a full analysis of the kinematic algebra for future work \cite{Cheung:future}.

The equations of motion in \Eq{eq:EOM} then become
\eq{
\frac{\delta S}{\delta W^{a}_A} &\simeq  \Box W^{a A} + f^{abc} F^{ABC} W_B^b W_C^c  =0 ,
}
where hereafter $\simeq$ will denote equality up to terms that vanish on the equations of motion either by \Eq{eq:dZ}  or by integration by parts, {\it e.g.}~$ \overset{\leftarrow}{\partial_\mu} Y + {\partial_\mu} Y + Y\overset{\rightarrow}{\partial_\mu}  =0$.
The field variations in \Eq{eq:deltaX}, \Eq{eq:deltaY}, \Eq{eq:deltaZ} become
\eq{
\delta W^A =   \left[ \begin{array}{c}
	\delta Z^\mu\\
	\delta Y \\
	\delta X_{\mu}  \end{array} \right] &
\simeq F^{ABC} \theta_B W_C ,
}
with the associated conservation equation,
\eq{
\partial J^{A} = \left[ \begin{array}{c}
	\partial J_X^{\mu} \\
	\partial J_Y \\
	\partial J_{Z\mu}  \end{array} \right]    \simeq - F^{ABC} W^a_B \overset{\leftrightarrow}{\Box}  W^a_C,
}
which is the kinematic analog of Eq.~\eqref{eq:color_Jacobi}, proving
\eq{
F^{DAE}F_E^{\;\;BC} + F^{DBE}F_E^{\;\;CA}+ F^{DCE}F_E^{\;\;AB} &\simeq 0,
}
which is precisely the kinematic Jacobi identity.

At the level of scattering amplitudes, these manipulations imply that all Feynman diagrams computed from \Eq{eq:action} will automatically satisfy Jacobi identities up to terms that vanish on the transverse condition in \Eq{eq:dZ}.

\section{Double Copy}

The double copy procedure maps \Eq{eq:colorAmp} to \Eq{eq:gravAmp} via a simple substitution of flavor with kinematics.  It is simple to show that the resulting double copy theory is the special Galileon~\cite{Cheung:future}, and in fact this is naturally anticipated by the CHY construction~\cite{Cachazo:2014xea}.

At the level of the action, the double copy is derived by mechanically dropping all flavor indices and doubling all kinematic structures in the interactions.  Since the action in \Eq{eq:action} is cubic, this is a trivial procedure.  To see how this works, we introduce the fields
\eq{
X_{\mu\bar \mu} ,\quad Y ,\quad Z^{\mu \bar \mu},
}
which have doubled index structure relative to the NLSM.
These new $XYZ$ fields couple via the action
\eq{
S		=  \int& \;\, Z^{\mu \bar\mu}  \Box X_{\mu \bar\mu}+    \frac{1}{2} \, Y  \Box Y \nonumber  \\
& +2 \left( Z^{\mu \bar\mu}Z^{\nu \bar\nu} X_{\mu\nu\bar{\mu}\bar{\nu}}  + Z^{\mu\bar\mu} (Y \overset{\leftrightarrow}{\partial_\mu} \overset{\leftrightarrow}{\partial_{\bar \mu}} Y  ) \right),
 \label{eq:sGalaction}
}
where we have defined an analog of Riemann curvature, $X_{\mu\nu\bar{\mu}\bar{\nu}}= \partial_\mu \partial_{\bar \mu} X_{\nu \bar\nu} +  \partial_\nu \partial_{\bar \nu} X_{\mu \bar\mu} - \partial_\mu \partial_{\bar \nu} X_{\nu \bar\mu} - \partial_\nu \partial_{\bar \mu} X_{\mu \bar\nu}$. Note that the barred and unbarred indices in \Eq{eq:sGalaction} are separately contracted, exhibiting the expected twofold Lorentz invariance of the double copy.

Tree amplitudes of the special Galileon are then given by \Eq{eq:amp} except where the ellipses denote doubly longitudinal polarizations of the $Z$ for which $\epsilon_{\mu\bar\mu} = i k_\mu k_{\bar\mu}$.  It would be interesting to understand how this construction relates to the Galileon as the longitudinal mode of massive gravity \cite{Nicolis:2008in}.

\section{Infrared Structure}

Lastly, we turn to infrared properties.
As the momentum $p$ of a particle is taken to be soft, amplitudes in the NLSM and the special Galileon scale as ${\cal O}(p)$ \cite{Adler:1964um} and ${\cal O}(p^3)$ \cite{Cheung:2014dqa, Hinterbichler:2015pqa}, respectively.  Remarkably,  these properties dictate virtually everything about these theories \cite{Cheung:2014dqa, Cheung:2016drk}, and can be leveraged to derive recursion relations for their amplitudes \cite{Cheung:2015ota}.  While this soft behavior usually obscured at the level of the action, the ${\cal O}(p)$ scaling of the NLSM and ${\cal O}(p^2)$ scaling of the special Galileon have a simple explanation in our formulation.  

In particular,  consider the soft limit of a Nambu-Goldstone boson, taken here to be a longitudinal $Z$ of the NLSM amplitude in \Eq{eq:amp}.
Since $Z$ enters with a derivative, the corresponding kinematic numerator trivially scales as ${\cal O}(p)$.  However, the hard leg from which $Z$ is emitted enters with a nearly on-shell propagator with ${\cal O}(p^{-1})$, so the net scaling of the amplitude is ${\cal O}(1)$.  Now observe from \Eq{eq:actionexpand} that cubic interactions of $Z$ take the form of gauge interactions modulo terms that vanish for longitudinal $Z$ components.  Although the action lacks the requisite  quartic interactions needed for a genuine $Z$ gauge symmetry, the soft $Z$ limit is dictated solely by cubic interactions. In turn, the $Z$ soft limit obeys the usual Weinberg soft theorems for gauge bosons \cite{Weinberg:1965nx}, dropping contributions from lower point amplitudes with a longitudinal $Z$ since they are odd and hence vanish by the underlying parity of the NLSM.  Gauge invariance then implies that the amplitude for soft longitudinal $Z$ emission is zero, eliminating the leading ${\cal O}(1)$ contribution but leaving the residual ${\cal O}(p)$ scaling of the NLSM.  This cancellation can be verified via Feynman diagrams.
Similarly, the ${\cal O}(p)$ contribution of the special Galileon vanishes by the Weinberg soft graviton theorem, however the further cancellation of ${\cal O}(p^2)$ terms is not obvious.

Remarkably, the leading nontrivial soft behavior of NLSM amplitudes is actually characterized by an underlying extended theory~\cite{Cachazo:2016njl}.  We can accommodate the structure by promoting $Y$ to a biadjoint field with the additional cubic coupling,
$
	 f^{abc}  f^{\bar a \bar b \bar c} Y^{a \bar a} Y^{b \bar b} Y^{c \bar c}$,
which preserves all the Jacobi identities of the full action.  We have verified that this modification reproduces the soft theorem in \cite{Cachazo:2016njl} up to ten-particle scattering.

\vspace*{-0.3cm}

\section{Conclusions}

In summary, we have reformulated the NLSM and special Galileon as theories of purely cubic interactions.  At the expense of explicit Bose symmetry and parity of the Nambu-Goldstone bosons, these representations exhibit several elegant properties.  In particular, they manifest flavor-kinematics duality as a symmetry, trivialize the double copy structure, and explain the vanishing soft behavior of amplitudes via the Weinberg soft theorem.

\medskip

\vspace*{0cm}

\noindent {\it Acknowledgments}: C.C. and C.-H.S. are supported by a Sloan Research Fellowship and a DOE Early Career Award under Grant No. DE-SC0010255. 

\bibliography{BCJ_NLSM}

\end{document}